\documentclass[12pt]{report}

\usepackage{latexsym}
\usepackage{bigints}
\usepackage[usenames]{color}
\usepackage{amssymb}
\usepackage{graphicx}
\usepackage{amssymb}
\usepackage{amscd}
\usepackage{eepic}
\usepackage{lmodern}
\usepackage{times}
\usepackage{cite}
\usepackage[french]{babel}
\usepackage[linktocpage, colorlinks=true,citecolor=blue,linkcolor=blue,urlcolor=blue]{hyperref}
\usepackage{color}
\usepackage{float}
\usepackage{amsmath, xspace}
\usepackage{amsthm}
\usepackage{amsfonts}
\usepackage{latexsym}
\usepackage{enumerate}
\RequirePackage{mathrsfs}
\RequirePackage[sc]{mathpazo}
\usepackage{palatino}
\newcommand{\be}{\begin{equation}}
\newcommand{\ee}{\end{equation}}
\newcommand{\bea}{\begin{eqnarray}}
\newcommand{\eea}{\end{eqnarray}}

\def \x {{\mathbf x}}
\def \d {\mathrm d}
\newcommand\bra[1]{\left\langle#1\right|}
\newcommand\ket[1]{\left|#1\right\rangle}
\newcommand\braket[2]{\left\langle#1|#2 \right\rangle}
\begin{document}
\title{Introduction aux m\'{e}thodes d'int\'{e}grale de chemin et applications \\
\bigskip
\small{Notes de cours  \\Master: Physique math\'{e}matique (2010-2012), \\ Centre de Physique Math\'{e}matique.}}

\author{Nour-Eddine Fahssi\\
\small \texttt{fahssi@fstm.ac.ma}\\}
\date{}
\maketitle
\begin{abstract}
Ces notes sont issues du cours que j'ai dispens\'{e} aux \'{e}tudiants de premi\`{e}re ann\'{e}e
inscrits \`{a} la facult\'{e} des sciences de l'universit\'{e} Mohammed V pour le master ``physique math\'{e}matique''. Elles sont bas\'{e}es pour l'essentiel sur les sources \'{e}num\'{e}r\'{e}es dans la bibliographie.
\end{abstract}
\tableofcontents
\chapter{Introduction}

Classiquement, il existe deux types de syst\`{e}mes dynamiques. Il y a d'abord le mouvement d'une particule ou d'un corps rigide avec un nombre fini de degr\'{e}s de libert\'{e}, qui peut \^{e}tre d\'{e}crit par un nombre fini de coordonn\'{e}es. Et puis, il y a des syst\`{e}mes physiques pour lesquels l'ensemble des degr\'{e}s de libert\'{e} est non d\'{e}nombrable. Ces syst\`{e}mes sont d\'{e}crits par des champs. Des exemples courants de champs classiques sont les champs \'{e}lectromagn\'{e}tiques d\'{e}crits par $E(\x,t)$ et $B (\x,t)$ ou de fa\c{c}on \'{e}quivalente par les potentiels ($\phi(\x,t), A(\x,t)$). De m\^{e}me, le mouvement d'une corde unidimensionnelle est \'{e}galement d\'{e}crite par un champ $\phi (\x,t)$. Ainsi, alors que les coordonn\'{e}es d'une particule ne d\'{e}pendent que du temps, les champs varient continuellement en fonction de certaines variables de l'espace. Par cons\'{e}quent, une th\'{e}orie d\'{e}crite par des champs est g\'{e}n\'{e}ralement vue comme une th\'{e}orie des champs \`{a} $D+1$ dimensions, o\`{u} $D$ repr\'{e}sente le nombre de dimensions spatiales. Par exemple, une th\'{e}orie d\'{e}crivant les d\'{e}placements de la corde unidimensionnelle constituerait une th\'{e}orie des champs de dimensions $1+1$ alors que la th\'{e}orie des \'{e}quations de Maxwell, plus familiers, peut \^{e}tre consid\'{e}r\'{e}e comme une th\'{e}orie de champs \`{a} $3+1$ dimensions. Dans cette perspective, il est alors clair que la th\'{e}orie d\'{e}crivant le mouvement d'une particule peut \^{e}tre particuli\`{e}rement consid\'{e}r\'{e}e comme une th\'{e}orie de champs \`{a} $0+1$ dimensions.

L'int\'{e}grale  de chemin est   une g\'{e}n\'{e}ralisation \`{a} un nombre infini de variables,  repr\'{e}sent\'{e} par  des chemins, des int\'{e}grales ordinaires. Elle v\'{e}rifie les m\^{e}mes propri\'{e}t\'{e}s alg\'{e}briques de celle-ci,  mais  pr\'{e}sente en plus des propri\'{e}t\'{e}s nouvelles.

La puissance des m\'{e}thodes de l'int\'{e}grale de chemin r\'{e}side dans le fait qu'elle met, tr\`{e}s explicitement, en correspondance les m\'{e}caniques
classique   et   quantique.   Les quantit\'{e}s  physiques   s'obtiennent en   moyennant
sur   tous   les   chemins possibles.   La caract\'{e}ristique int\'{e}ressante de l'int\'{e}grale de chemin est que dans   la   limite semi-classique  $\hbar \rightarrow 0$   les chemins  dominant   l'int\'{e}grale se trouvent   dans   un   voisinage du   chemin ``classique'', ce qui permet de bien comprendre le principe de correspondance.

La formulation de la m\'{e}canique quantique bas\'{e}e sur   l'int\'{e}grale de
chemin est   plus compliqu\'{e}e du point   de vue math\'{e}matique. Cependant, elle
est bien adapt\'{e}e \`{a} l'\'{e}tude de syst\`{e}mes \`{a} grand nombre de degr\'{e}s de libert\'{e}
o\`{u} un   formalisme   bas\'{e} sur l'\'{e}quation de Schr\"{o}dinger perd toute son utilit\'{e}.
Elle permet ainsi une analyse ais\'{e}e de  la   th\'{e}orie   quantique   des   champs   et de   la   m\'{e}canique
statistique.

\section{Notes historiques} En 1933, Dirac a observ\'{e} que l'action joue un r\^{o}le central dans la m\'{e}canique classique (il consid\`{e}re que la formulation lagrangienne de la m\'{e}canique classique est plus fondamentale que la formulation hamiltonienne), mais qu'il ne semblait avoir aucun r\^{o}le important dans la m\'{e}canique quantique. Il a ensuite avanc\'{e} que cette situation pourrait \^{e}tre corrig\'{e}e si le propagateur de la m\'{e}canique quantique ``correspondait \`{a}'' $\exp i S/\hbar$ o\`{u} $S$ est l'action classique \'{e}valu\'{e}es le long du chemin classique.

En 1948, Feynman a d\'{e}velopp\'{e} la suggestion de Dirac, et a r\'{e}ussi \`{a} d\'{e}river une troisi\`{e}me formulation de la m\'{e}canique quantique, bas\'{e}e sur le fait que le propagateur peut \^{e}tre \'{e}crit comme une somme sur tous les chemins possibles (pas seulement les classiques) reliant le point initial et le point final du mouvement d'une particule.  Ainsi, alors que Dirac n'a consid\'{e}r\'{e} que le chemin classique, Feynman a montr\'{e} que tous les chemins contribuent. Pour l'anecdote, l'article original de Feynman qui a jet\'{e} les bases du sujet a \'{e}t\'{e} rejet\'{e} par Physical Review!

\section{Notations}
Dans ce cours, nous allons consid\'{e}rer les th\'{e}ories non-relativistes et relativistes. Pour le cas relativiste nous allons utiliser la convention de Bjorken-Drell : les coordonn\'{e}es contravariantes  sont suppos\'{e}es \^{e}tre
\be \label{1} x^\mu=(t,\x),    \quad     \mu=0,1,2,3 \, ,\ee
tandis que les coordonn\'{e}es covariantes sont \be \label{2} x_\mu=\eta_{\mu \nu}x^\nu=(t,-\x) .\ee
Tout au long de ce cours, nous posons $c=1$. La m\'{e}trique covariante est donc diagonale : \be \eta_{\mu \nu}=(+,-,-,-)\ee la m\'{e}trique inverse ou contravariante a la m\^{e}me forme  \be \eta^{\mu \nu}=(+,-,-,-) .\ee l'\'{e}l\'{e}ment de longueur invariant est \be x^2=x_\mu x^\mu = \eta_{\mu \nu}x^\mu x^\nu= \eta^{\mu \nu}x_\mu x_\nu=t^2-\x^2 .\ee
Les gradients sont \'{e}galement obtenus \`{a} partir de~\eqref{1} et~\eqref{2}: \bea \partial_\mu &=& \frac{\partial}{\partial x^\mu }=\left(\frac{\partial}{\partial t},\nabla\right),\\ \partial ^\mu &=& \frac{\partial}{\partial x_\mu}=\left(\frac{\partial}{\partial t},-\nabla\right),\eea le D'Alembertien prend donc la forme \be \Box = \partial^\mu \partial_\mu = \eta^{\mu \nu}\partial_\mu \partial_ \nu =\frac{\partial^2}{\partial t^2}-\nabla^2 .\ee

\section{Fonctionnelles}
Dans l'\'{e}tude de n'importe quel syst\`{e}me dynamique, on a souvent affaire avec des fonctions dont la ``variable'' est une fonction. Ces fonctions compos\'{e}es s'appellent des fonctionnelles. Si nous consid\'{e}rons le mouvement d'une particule plong\'{e}e dans un potentiel \`{a} une dimension, alors le Lagrangien est donn\'{e} par \be  L(x, \dot{x})=\frac{1}{2}m \dot{x}^2-V(x), \ee o\`{u} $x(t)$ et $\dot{x}(t)$ sont la coordonn\'{e}e et la vitesse de la particule. La fonctionnelle la plus simple \`{a} laquelle on peut penser en pr\'{e}sence d'un lagrangien est l'action d\'{e}finie par \be \label{action} S[x]=\int_{t_i}^{t_f} \d t \, L(x,\dot{x})\ee
Notons que, contrairement \`{a} une fonction dont la valeur d\'{e}pend d'un point particulier dans l'espace de coordonn\'{e}es, la valeur de l'action d\'{e}pend de la trajectoire tout le long de laquelle l'int\'{e}gration est effectu\'{e}e. Pour les diff\'{e}rents chemins reliant le point initial $x(t_i)$ au point final $x(t_f)$, les valeurs de la fonctionnelle action seront diff\'{e}rentes.

Ainsi, une fonctionnelle a la forme g\'{e}n\'{e}rique suivante \be F[f]=\int \d x \, F(f(x)) \ee

La notion de d\'{e}riv\'{e}e peut \^{e}tre \'{e}tendue au cas de fonctionnelles de fa\c{c}on naturelle \`{a} travers la notion de fonctions g\'{e}n\'{e}ralis\'{e}es (ou distributions). Ainsi, on d\'{e}finit la d\'{e}riv\'{e}e fonctionnelle ou \emph{la d\'{e}riv\'{e}e de G\^{a}teaux} par la fonctionnelle lin\'{e}aire
\be F'[v]=\frac{\d}{\d \epsilon} F[f+\epsilon v]\Big|_{\epsilon=0}=\int \d x \frac{\delta F[f]}{\delta f(x)} v(x).\ee la quantit\'{e} $\frac{\delta F[f]}{\delta f(x)}$ s'appelle d\'{e}riv\'{e}e fonctionnelle de $F[f]$. Elle est d\'{e}finie par \be \frac{\delta F[f(x)]}{\delta f(y)}=\lim _{\epsilon \rightarrow 0}\frac{F(f(x)+\epsilon \delta (x-y))-F(f(x))}{\epsilon}\ee Il s'ensuit de cette d\'{e}finition que \be \frac{\delta f(x)}{\delta f(y)}=\delta(x-y).\ee o\`{u}  $\delta(x)$ est la fonction (plus exactement distribution) de Dirac.

La d\'{e}riv\'{e}e fonctionnelle satisfait toutes les propri\'{e}t\'{e}s de la d\'{e}riv\'{e}e, notamment
\bea \nonumber \frac{\delta}{\delta f(x)}(F_1[f]+F_2[f])&=&\frac{\delta F_1[f]}{\delta f(x)}+\frac{\delta F_2[f]}{\delta f(x)}\\
\frac{\delta}{\delta f(x)}(F_1[f]F_2[f])&=&\frac{\delta F_1[f]}{\delta f(x)}F_2[f]+F_1[f]\frac{\delta F_2[f]}{\delta f(x)}\eea En outre, une fonctionnelle peut \^{e}tre d\'{e}velopp\'{e}e en s\'{e}rie de Taylor comme suit:
\be \label{taylor}F[f]=\int \d x P_0(x) +\int \d x_1 \d x_2 P_1(x_1,x_2) f(x_2) + \int \d x_1 \d x_2 \d x_3 P_2(x_1,x_2,x_3) f(x_2) f(x_3) + \ldots ,\ee o\`{u} \bea \nonumber P_0(x) &=& F(f(x)) \big |_{f(x)=0}\, , \\
P_1(x_1,x_2)&=&\frac{\delta F(f(x_1))}{\delta f(x_2)}\Big |_{f(x)=0}\, ,\\
\nonumber P_2(x_1,x_2,x_3)&=&\frac{1}{2!}\frac{\delta^2 F(f(x_1))}{\delta f(x_2) \delta f(x_3)}\Big |_{f(x)=0}\, ,
\eea etc.

\subsection{Examples}
\textbf{(i)} Soit $F(f)=f(x)^n$, o\`{u} $n$ est un entier positif.
\be F[f]=\int \d y F(f(y))=\int \d y f(y)^n \, .\ee Alors, \bea \nonumber \frac{\delta F[f(y)]}{\delta f(x)} &=&\lim _{\epsilon \rightarrow 0}\frac{F(f(y)+\epsilon \delta (y-x))-F(f(y))}{\epsilon}\\
\nonumber &=&\lim _{\epsilon \rightarrow 0}\frac{(f(y)+\epsilon \delta (y-x))^n-f(y)^n}{\epsilon}\\
\nonumber &=& \lim _{\epsilon \rightarrow 0}\frac{f(y)^n+ n\epsilon f(y)^{n-1}\delta (y-x) +O(\epsilon^2)-f(y)^n}{\epsilon} \\
&=& nf(y)^{n-1} \delta(y-x).\eea
Par cons\'{e}quent, nous obtenons \bea \nonumber \frac{\delta F[f]}{\delta f(x)} &=& \int \d y \frac{\delta F[f(y)]}{\delta f(x)}\\
\nonumber &=& \int \d y \,n f(y)^{n-1} \delta(y-x)\\
 &=& n f(x)^{n-1}\eea

\textbf{(ii)} Consid\'{e}rons ensuite l'action~\eqref{action} \be S[x(t)] = \int_{t_i}^{t_f} \d t' \, L(x(t'),\dot{x}(t'))\, ,\ee avec \bea \nonumber L(x(t),\dot{x}(t))&=&\frac{1}{2}m \dot{x}(t)^2-V(x(t))\\
&=&T(\dot{x}(t))-V(x(t)).\eea
Nous obtenons sans difficult\'{e} \bea \nonumber \frac{\delta V(x(t'))}{\delta x(t)}&=& \lim _{\epsilon \rightarrow 0}\frac{V(x(t')+\epsilon \delta (t'-t))-V(x(t'))}{\epsilon}\\
&=& V'(x(t')) \delta(t'-t),\eea o\`{u} $V'(x(t'))=\partial V(x(t')) / \partial x(t')$. De m\^{e}me,
 \bea \nonumber \frac{\delta T(\dot{x}(t'))}{\delta x(t)}&=& \lim _{\epsilon \rightarrow 0}\frac{T(\dot{x}(t')+\epsilon \frac{\d}{\d t'}\delta (t'-t))-T(\dot{x}(t'))}{\epsilon}\\
&=& m \dot{x}(t') \frac{\d}{\d t'}\delta(t'-t).\eea Il est clair maintenant que \bea \nonumber \frac{\delta L(x(t'),\dot{x}(t'))}{\delta x(t)} &=& \frac{\delta T(\dot{x}(t')-V(x(t')))}{\delta x(t)}\\
&=& m \dot{x}(t') \frac{\d}{\d t'}\delta(t'-t)-V'(x(t'))\delta(t'-t).\eea
Nous obtenons donc pour $t_i \leq t \leq t_f$ \bea  \nonumber \frac{\delta S[x(t)]}{\delta x(t)}&=& \int_{t_i}^{t_f} \d t' \, \frac{\delta L(x(t'),\dot{x}(t'))}{\delta x(t)}\\
\nonumber &=&\int_{t_i}^{t_f} \d t' \, \left(m \dot{x}(t') \frac{\d}{\d t'}\delta(t'-t)-V'(x(t'))\delta(t'-t)\right) \\
\nonumber &=& -m \ddot{x}(t)-V'(x(t))\\
&=& -\frac{\d}{\d t}\frac{\partial L(x(t),\dot{x}(t))}{\partial \dot{x}(t)}+\frac{\partial L(x(t),\dot{x}(t))}{\partial x(t)}\eea
Ce dernier membre rappelle l'\'{e}quation d'Euler-Lagrange. En fait, on peut noter que \be \frac{\delta S[x(t)]}{\delta x(t)} = -\frac{\d}{\d t}\frac{\partial L(x(t),\dot{x}(t))}{\partial \dot{x}(t)}+\frac{\partial L(x(t),\dot{x}(t))}{\partial x(t)} = 0\ee n'est autre que l'\'{e}quation d'Euler-Lagrange \'{e}crite comme extremum de la fonctionnelle action. Ceci constitue exactement le principe de la moindre action exprim\'{e}e dans le langage des fonctionnelles.
\section{Un bref rappel de m\'{e}canique quantique}
Dans cette section, nous allons d\'{e}crire tr\`{e}s bri\`{e}vement les \'{e}l\'{e}ments essentiels de la m\'{e}canique quantique. L'approche conventionnelle \`{a} la m\'{e}canique quantique commence par la formulation hamiltonienne de la m\'{e}canique classique en consid\'{e}rant que les observables sont des op\'{e}rateurs qui ne commutent pas. La dynamique, dans ce cas, est donn\'{e} par l'\'{e}quation de Schr\"{o}dinger d\'{e}pendante du temps \be i \hbar {\partial |\psi (t) \rangle \over \partial t}=H |\psi (t) \rangle ,\ee o\`{u} $H$ est l'op\'{e}rateur hamiltonien du syst\`{e}me. De fa\c{c}on \'{e}quivalente, la fonction d'onde d'une particule, dans le cas uni-dimensionnel, satisfait
\bea \nonumber  i \hbar  {\partial \psi (x,t) \over \partial t} &=& H(x) \psi (x,t)\\
&=& \left(-\frac{\hbar^2}{2m}\frac{\partial^2}{\partial x^2}+V(x)\right) \psi (x,t), \eea
o\`{u} l'on a pos\'{e} \be \psi(x,t) = \bra{ x}\psi(t) \rangle ,\ee $|x\rangle$ d\'{e}signent les \'{e}tats de base dans la representation de coordonn\'{e}es. Cela d\'{e}finit alors l'\'{e}volution dans le temps du syst\`{e}me.

L'object principal de la r\'{e}solution de l'\'{e}quation de Schr\"{o}dinger est la d\'{e}termination de l'op\'{e}rateur d'\'{e}volution qui g\'{e}n\`{e}re la translation de ce syst\`{e}me dans le temps. A savoir, l'op\'{e}rateur d'\'{e}volution transformant l'\'{e}tat quantique \`{a} un temps pass\'{e} $t_2$ \`{a} l'\'{e}tat quantique \`{a} un temps avenir $t_1$ : \be \label{evol}\ |\psi (t_1) \rangle = U(t_1,t_2) |\psi (t_2) \rangle .\ee
Clairement, pour un hamiltonien ind\'{e}pendant du temps, on voit, \`{a} partir de l'\'{e}quation de Schr\"{o}dinger \be U(t_1,t_2)=\theta(t_1-t_2) e^{-\frac{i}{\hbar}(t_1-t_2) H}\ee o\`{u} $\theta$ est la fonction \'{e}chelon de Heaviside.

Il est \'{e}vident que l'op\'{e}rateur d'\'{e}volution temporelle n'est autre que la fonction de Green pour l'\'{e}quation de Schr\"{o}dinger v\'{e}rifiant l'\'{e}quation diff\'{e}rentielle : \be\left(i \hbar {\partial \over \partial t_1}-H\right) U(t_1,t_2) = i \hbar \delta (t_1-t_2). \ee (Note. la d\'{e}riv\'{e}e au sens des distributions de $\theta$ est \'{e}gale \`{a} la distribution de Dirac $\delta$). Pour d\'{e}terminer l'op\'{e}rateur $U$, il faut calculer ses \'{e}l\'{e}ments de matrices dans une base donn\'{e}e. Donc, par exemple, dans la base des vecteurs propres de l'op\'{e}rateur position $X$ : $\{| x \rangle\}$ d\'{e}finie par \be X | x \rangle = x | x \rangle ,\ee nous pouvons \'{e}crire
\be \bra{x_1}U(t_1,t_2)\ket{x_2} = U(t_1,x_1 ; t_2,x_2).\ee
Si on sait calculer compl\`{e}tement la fonction $U(t_1,x_1; t_2,x_2)$, alors l'\'{e}volution dans le temps de la fonction d'onde $\psi$ peut \^{e}tre exprim\'{e}e par \bea \nonumber \psi (x_1,t_1) &=& \bra{ x_1} U(t_1,t_2) \ket{ \psi (t_2)} \\
\nonumber &=& \int \d x_2  \, \bra{ x_1}U(t_1,t_2) \ket{x_2}  \langle x_2 | \psi (t_2)\rangle \\
 &=& \int \d x_2   \, U(t_1,x_1 ; t_2,x_2)  \psi (x_2 , t_2). \eea Dans ce calcul, nous avons utilis\'{e} la relation de fermeture : \be \int \d x \, |x\rangle \langle x|=\mathbf{1}.\ee

 Le sch\'{e}ma adopt\'{e} dans la discussion pr\'{e}c\'{e}dente est celui de Schr\"{o}dinger, o\`{u} les \'{e}tats quantiques $| \psi (t)\rangle $ sont d\'{e}pendants du temps tandis que les op\'{e}rateurs ne le sont pas.
 \begin{center}
 \begin{tabular}{l|l|l}
   Sch\'{e}ma &Schr\"{o}dinger & Heisenberg \\
    \hline \hline
   \'{E}tats  & d\'{e}p de t& ind\'{e}p de t\\
   \hline
   Op\'{e}rateurs & ind\'{e}p de t & d\'{e}p de t\\
   \hline
 \end{tabular}
 \end{center}
 D'autre part, dans le sch\'{e}ma de Heisenberg, o\`{u} les \'{e}tats sont ind\'{e}pendants du temps, nous pouvons \'{e}crire par~\eqref{evol} \be | \psi \rangle_H = | \psi (t=0)\rangle_S = | \psi (t=0)\rangle = e^{\frac{i}{\hbar}t H}| \psi (t)\rangle = e^{\frac{i}{\hbar}t H}| \psi (t)\rangle_S.\ee

 Dans ce sch\'{e}ma, l'op\'{e}rateur position est li\'{e} \`{a} celui dans le sch\'{e}ma de Schr\"{o}dinger par \be X_H(t)=e^{\frac{i}{\hbar}t H} X e^{-\frac{i}{\hbar}t H}.\ee Les \'{e}tats propres de ces op\'{e}rateurs satisfont \be X_H(t) \ket{x,t}_H = x \ket{x,t}_H, \ee et par cons\'{e}quent \be \ket{x,t}_H=e^{\frac{i}{\hbar}t H}\ket{x}.\ee Pour $t_1 > t_2$, il est clair que
 \bea \nonumber _H\braket{x_1,t_1}{x_2,t_2}_H &=& \bra{x_1}e^{-\frac{i}{\hbar}t_1 H} e^{\frac{i}{\hbar}t_2 H}\ket{x_2}\\
\nonumber &=& \bra{x_1}e^{-\frac{i}{\hbar}(t_1-t_2) H}\ket{x_2}\\
\nonumber &=& \bra{x_1}U(t_1,t_2)\ket{x_2}\\
 &=& U(t_1,x_1 ; t_2,x_2)\eea Cela montre que les \'{e}l\'{e}ments de matrice de l'op\'{e}rateur d'\'{e}volution temporelle n'est autre que l'amplitude de transition entre les \'{e}tats de base dans le sch\'{e}ma de Heisenberg.

\chapter{Int\'{e}grale de chemin en m\'{e}canique quantique}
\section{Formules de base de la m\'{e}canique quantique} Avant de d\'{e}terminer la repr\'{e}sentation par l'int\'{e}grale de chemin de l'amplitude de transition $U(t_f,x_f;t_i,x_i)$, r\'{e}capitulons quelques formules de base. Consid\'{e}rons, pour simplifier, un syst\`{e}me uni-dimensionnel. Les \'{e}tats propres de l'op\'{e}rateur position $X$ forment une base orthonorm\'{e}e : \[\braket{x}{x'}=\delta(x-x'),\]
\be \int \d x \, \ket{x}\!\bra{x} = \mathbf{1}.\ee Les \'{e}tats propres de l'op\'{e}rateur impulsion $P$ forment \'{e}galement une base orthonormale : \be P\ket{p}=p\ket{p},\ee et \[\braket{p}{p'}=\delta(p-p'),\]
\be \int \d p \, \ket{p}\!\bra{p} = \mathbf{1}.\ee Le produit scalaire $\braket{x}{p}$ donne les \'{e}l\'{e}ments de matrice de l'op\'{e}rateur de passage entre les deux bases $\{\ket{x}\}$ et $\{\ket{p}\}$. En fait, on peut calculer \be \braket{p}{x} = \frac{1}{\sqrt{2 \pi \hbar}} e^{-\frac{i}{\hbar}p x}=\braket{x}{p}^*.\ee Ces relations d\'{e}finissent des transformations de Fourier. Plus pr\'{e}cis\'{e}ment, \`{a} l'aide de la relation de fermeture, on peut \'{e}crire la transformation de Fourier d'une fonction comme
\bea f(x)= \nonumber \braket{x}{f}&=& \int \d p \, \braket{x}{p} \braket{p}{f}\\
\nonumber &=& \frac{1}{\sqrt{2\pi \hbar}}\int \d p \, e^{\frac{i}{\hbar}p x}f(p)\\
&=&\frac{1}{\sqrt{2\pi}}\int \d k \, e^{i k x}\widetilde{f}(k)\eea o\`{u} \bea \nonumber \widetilde{f}(k)&=&\sqrt{\hbar} f(p)\\
\nonumber &=& \frac{\sqrt{\hbar}}{\sqrt{2 \pi \hbar}}\int \d x \, e^{-\frac{i}{\hbar}p x} f(x)\\
&=&\frac{1}{\sqrt{2 \pi}} \int \d x \,  e^{-i k x}f(x).\eea Ici, $k = p/\hbar$ peut \^{e}tre vu comme le nombre d'onde.

Dans le sch\'{e}ma de Heisenberg, les \'{e}tats quantiques v\'{e}rifient \bea \nonumber _H \braket{x,t}{x',t}_H &=& \bra{x} e^{-\frac{i}{\hbar}t H}e^{\frac{i}{\hbar}t H}\ket{x'}\\
&=& \braket{x}{x'} = \delta(x-x'), \eea
et \bea \label{ferm}\nonumber\int \d x \, \ket{x,t}_H \bra{x,t}_H &=& \int \d x \, e^{\frac{i}{\hbar}t H} \ket{x} \bra{x} e^{-\frac{i}{\hbar}t H} \\
\nonumber &=& e^{\frac{i}{\hbar}t H} \int \d x \, \ket{x} \bra{x} e^{-\frac{i}{\hbar}t H}\\
&=& e^{\frac{i}{\hbar}t H}\mathbf{1} e^{-\frac{i}{\hbar}t H}= \mathbf{1}.\eea Il faut noter que l'orthonormalit\'{e} ainsi que la relation de fermeture ne sont valables dans le sch\'{e}ma de Heisenberg que pour des temps $t$ \'{e}gaux.

\section{Ordre des op\'{e}rateurs en m\'{e}canique quantique}
Les op\'{e}rateurs en m\'{e}canique quantique ne commutent pas. Par cons\'{e}quent, le hamiltonien classique est sens\'{e} ``devenir'' un op\'{e}rateur quantique : $$ H(x,p) \longrightarrow H(X,P).$$ Cependant, ce passage ne sp\'{e}cifie pas l'ordre d'\'{e}criture de l'analogue quantique d'un terme comme $xp$ par exemple. Est-ce que c'est $XP$ ou, plut\^{o}t $PX$. Comme cet ordre n'importe pas dans la m\'{e}canique classique, il est crucial en m\'{e}canique quantique, et, \`{a} priori, il n'est pas clair qu'un ordre donn\'{e} doit correspondre \`{a} la th\'{e}orie quantique. c'est le probl\`{e}me de l'ordre des op\'{e}rateurs et, malheureusement, il n'existe aucun principe bien d\'{e}fini dans ce sens. Il y a cependant quelques conventions que l'on adopte en g\'{e}n\'{e}ral. Dans l'ordre ``normal'', on ordonne les produits des op\'{e}rateurs $X$ et $P$ tels que les impulsions se mettent \`{a} gauche des coordonn\'{e}es. Ainsi \[\begin{array}{lcl}
                                                                                                                                             xp & \stackrel{\mbox{\tiny O.N}}\longrightarrow & PX \\
                                                                                                                                             px & \stackrel{\mbox{\tiny O.N}}\longrightarrow & PX \\
                                                                                                                                             x^2 p & \stackrel{\mbox{\tiny O.N}}\longrightarrow & P X^2 \\
                                                                                                                                             xpx & \stackrel{\mbox{\tiny O.N}}\longrightarrow & P X^2
                                                                                                                                           \end{array}
\] etc. Cependant, il existe une autre convention plus satisfaisante et qui est plus largement adopt\'{e}e. Il s'agit de l'ordre de Weyl qui consiste \`{a} symm\'{e}triser les produits d'op\'{e}rateurs dans toutes les combinaisons possibles avec le m\^{e}me coefficient de pond\'{e}ration. Ainsi,
 \[\begin{array}{lcl}
                                                                                                                                             xp & \stackrel{\mbox{\tiny O.W}}\longrightarrow & \frac{1}{2}(XP+PX) \\
                                                                                                                                             px & \stackrel{\mbox{\tiny O.W}}\longrightarrow & \frac{1}{2}(XP+PX) \\
                                                                                                                                             x^2 p & \stackrel{\mbox{\tiny O.W}}\longrightarrow & \frac{1}{3}(X^2 P+XPX+P X^2) \\
                                                                                                                                             xpx & \stackrel{\mbox{\tiny O.W}}\longrightarrow & \frac{1}{3}(X^2 P+XPX+P X^2)
                                                                                                                                           \end{array}
\] etc. Pour l'ordre normal, c'est un exercice facile de montrer que pour tout hamiltonien quantique obtenu \`{a} partir d'un hamiltonien classique $H(x,p)$ on a \bea\nonumber \bra{x'}:H:_{\, \mbox {\tiny O.N}}\ket{x}&=&\int \d p \, \braket{x'}{p}\bra{p}:H:_{\, \mbox{\tiny O.N}}\ket{x}\\
&=&\int \frac{\d p}{2 \pi \hbar} \, e^{-\frac{i}{\hbar}p (x-x')} \, H(x,p).\eea
On peut de m\^{e}me montrer que sous l'ordre de Weyl, cette relation s'\'{e}crit \be \label{weyl}  \bra{x'}:H:_{\, \mbox {\tiny W}}\ket{x}=\int \frac{\d p}{2 \pi \hbar} \, e^{-\frac{i}{\hbar}p (x-x')} \, H\left(\frac{x+x'}{2},p\right).\ee Comme on le voit, les \'{e}l\'{e}ments de matrice du hamiltonien sous l'ordre de Weyl conduit \`{a} ce qu'on appelle la prescription du point milieu (mid-point) et c'est ce que nous allons utiliser dans toutes nos discussions.

\section{Calcul de l'amplitude par l'int\'{e}grale de chemin}
Nous sommes maintenant pr\^{e}ts pour le calcul de l'amplitude de transition (ou propagateur). Rappelons que dans le sch\'{e}ma de Heisenberg, pour $t_f > t_i$, nous avons
\[U(t_f,x_f;t_i,x_i) = {~}_H \braket{x_f,t_f}{x_i,t_i}_H.\]
Divisons l'intervalle entre les temps initial et final en $N$ segments infinit\'{e}simaux de longueur $\epsilon$. Soit \be \epsilon = {t_f-t_i \over N}.\ee En d'autres termes, pour la simplicit\'{e}, nous allons discr\'{e}tiser l'intervalle de temps et, \`{a} la fin de nos calculs, nous passerons \`{a} la limite $\epsilon \rightarrow 0$ et $N \rightarrow \infty$. Un instant interm\'{e}diaire $t_n$ peut donc s'\'{e}crire
\be t_n=t_i+n \epsilon, \quad n=1,2, \ldots , N-1.\ee

Utilisant la relation de fermeture (Eq.~\eqref{ferm}) pour tous les instants interm\'{e}diaires, nous obtenons \begin{multline}
  U(t_f,x_f;t_i,x_i) = \lim_{{\epsilon \rightarrow 0 \atop N\rightarrow \infty}} \int \d x_1 \cdots \d x_{N-1} \, _H \braket{x_f,t_f}{x_{N-1},t_{N-1}}_H \\ \times _H \braket{x_{N-1},t_{N-1}}{x_{N-2},t_{N-2}}_H \cdots _H\braket{x_1,t_1}{x_i,t_i}_H .\\
\end{multline}
Notons que tout produit scalaire dans cette formule admet la forme
\bea \nonumber {~}_H \braket{x_n,t_n}{x_{n-1},t_{n-1}}_H  &=& \bra{x_n} e^{-\frac{i}{\hbar}(t_n - t_{n-1}) H} \ket{x_{n-1}} \\
\nonumber &=& \bra{x_n} e^{-\frac{i}{\hbar}\epsilon H} \ket{x_{n-1}} \\
&=& \int \frac{\d p_n}{2 \pi \hbar} \, e^{\frac{i}{\hbar}p_n (x_n-x_{n-1})-\frac{i}{\hbar}\epsilon H\left(\frac{x_n+x_{n-1}}{2},p_n\right)}\eea
Ici nous avons utilis\'{e} la prescription du point milieu de l'ordre de Weyl~\eqref{weyl}.

Substituons cette forme du produit scalaire dans l'amplitude de transition, nous obtenons
\begin{multline}\label{PI} U(t_f,x_f;t_i,x_i) = \lim_{{\epsilon \rightarrow 0 \atop N\rightarrow \infty}} \int \d x_1 \cdots \d x_{N-1} \frac{\d p_1}{2 \pi \hbar} \cdots \frac{\d p_N}{2 \pi \hbar} \\ \times e^{\frac{i}{\hbar}\sum_{n=1}^N \left(p_n (x_n-x_{n-1})-\epsilon H\left(\frac{x_{n} + x_{n-1}}{2},p_n\right)\right)}
.\end{multline} o\`{u} l'on a pos\'{e} $x_0=x_i$ et $x_N=x_f$. C'est le forme la plus grossi\`{e}re de l'int\'{e}grale de chemin de Feynman et est d\'{e}finie
dans l'espace de phase du syst\`{e}me. Le facteur de phase dans le membre de droite de la derni\`{e}re \'{e}quation s'\'{e}crit dans la limite $\epsilon \rightarrow \infty$
\bea \lim_{{\epsilon \rightarrow 0 \atop N \rightarrow \infty}}\nonumber \frac{i}{\hbar}\epsilon \sum_{n=1}^N p_n \left(\frac{x_n-x_{n-1}}{\epsilon}\right)- H\left(\frac{x_{n} + x_{n-1}}{2},p_n\right) &=& \frac{i}{\hbar}\int_{t_i}^{t_f} \d t \, (p \dot{x} - H(x,p)) \\
&=& \frac{i}{\hbar}\int_{t_i}^{t_f} \d t \, L \; .\eea En d'autres termes, il est proportionnel \`{a} l'action.
\begin{figure}[htbp]
 \centering
  \includegraphics[width=8cm]{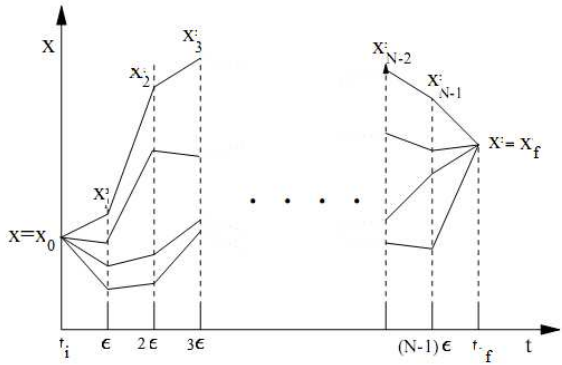}\\
\caption{L'amplitude comme somme sur tous les chemins $N$-segment\'{e}s.}
\end{figure} Pour obtenir la forme la plus famili\`{e}re de l'int\'{e}grale de chemin impliquant le Lagrangien dans l'espace de configuration ( c'est \`{a} dire une int\'{e}grale sur les variables $x$ uniquement), nous consid\'{e}rons la classe des Hamiltoniens qui sont quadratiques par rapport aux impulsions $p$. Choisissons \be H(x,p)=\frac{p^2}{2m}+V(x) . \ee Dans ce cas, nous avons par l'\'{e}quation~\eqref{PI}
\begin{multline}\label{PI1}U(t_f,x_f;t_i,x_i) = \lim_{{\epsilon \rightarrow 0 \atop N\rightarrow \infty}} \int \d x_1 \cdots \d x_{N-1} \frac{\d p_1}{2 \pi \hbar} \cdots \frac{\d p_N}{2 \pi \hbar} \\ \times e^{\frac{i \epsilon}{\hbar}\sum_{n=1}^N \left(p_n \left( \frac{x_n-x_{n-1}}{\epsilon}\right)-\frac{p_n^2}{2m} -V\left(\frac{x_{n} + x_{n-1}}{2}\right)\right)}
.\end{multline} Les int\'{e}grales sur les impulsions sont des int\'{e}grales gaussiennes que l'on peut facilement calculer. Utilisant la formule \be \int_0^\infty e^{- i \alpha t^2} \d t =\frac{1}{2}\sqrt{\frac{\pi}{\alpha }}e^{-\frac{i \pi}{4}},\ee on trouve \bea \nonumber \int \frac{\d p_n}{2 \pi \hbar} \, e^{-i \frac{\epsilon}{ \hbar} \left(\frac{p_n^2}{2m}-\frac{p_n (x_n-x_ {n-1})}{\epsilon}\right)} &=&\int \frac{\d p_n}{2 \pi \hbar} \, e^{ -\frac{i\epsilon}{2m \hbar}\left\{\left(p_n-\frac{m(x_n-x_{n-1})}{\epsilon}\right)^2-\left(\frac{m(x_n-x_{n-1})}{\epsilon}\right)^2\right\}} \\
&=& \left(\frac{m}{2 \pi i \hbar \epsilon}\right)^{1/2}\, \exp\left\{\frac{i m \epsilon}{2 \hbar}\left(\frac{x_n- x_ {n-1}}{\epsilon}\right)^2\right\}.\eea Rempla\c{c}ant en plus dans l'\'{e}quation~\eqref{PI1}, nous obtenons \bea \label{FPI}\nonumber U(t_f,x_f;t_i,x_i) &=& \lim_{{\epsilon \rightarrow 0\atop N \rightarrow \infty}}\left(\frac{m}{2 \pi i \hbar \epsilon}\right)^{N/2} \\ \nonumber && \times \int \d x_1 \cdots \d x_{N-1} \, e^{\frac{i \epsilon}{\hbar}\sum_{n=1}^N \left(\frac{m}{2}\left(\frac{x_n -x_{n-1}}{\epsilon}\right)^2-V\left(\frac{x_n +x_{n-1}}{2}\right)\right)}\\
\nonumber &=& A \int \mathscr{D}[x(t)] \, e^{\frac{i}{\hbar}\int_{t_i}^{t_f} \d t \left(\frac{1}{2}m \dot{x}^2 -V(x)\right)}\\
&=& A \int \mathscr{D}[x(t)] \, e^{\frac{i}{\hbar}S[x(t)]}, \eea o\`{u} $A$ est une constante ind\'{e}pendante de la dynamique du syst\`{e}me  et $S[x(t)]$ est l'action classique donn\'{e}e par~\eqref{action}. C'est l'int\'{e}grale de chemin de Feynman pour l'amplitude de transition en m\'{e}canique quantique.

L'int\'{e}grale de Feynman stipule que l'amplitude de transition entre un \'{e}tat initial et un \'{e}tat final est simplement la somme sur tous les chemins,  reliant les deux points, du facteur $\exp(i/\hbar \, S[x(t)])$. Cette somme est traduite par l'int\'{e}grale sur les coordonn\'{e}es spatiales avec une mesure d'int\'{e}gration donn\'{e}e par $\mathscr{D}[x(t)]$.

\paragraph{Exercices.} \begin{enumerate}
                         \item Equivalence de la m\'{e}thode d'int\'{e}grale de chemin avec l'\'{e}quation de Schr\"{o}dinger.
                         \item Montrer que l'amplitude de transition pour une particule libre s'\'{e}crit \be \label{libre}U(t_f,x_f;t_f,t_i)=\left(\frac{m}{2 \pi i \hbar (t_f-t_i)}\right)^{\frac{1}{2}}e^{\frac{i}{\hbar}\frac{m(x_f-x_i)^2}{2(t_f-t_i)}}.\ee
                         \begin{enumerate}
                           \item Montrer que ce propagateur est une solution de l'\'{e}quation de Schr\"{o}dinger.
                         \end{enumerate}
                       \end{enumerate}

\chapter{Int\'{e}grale de chemin pour l'oscillateur harmonique}
Comme exemple important d'application de l'int\'{e}grale de chemin, nous allons consid\'{e}rer l'oscillateur harmonique uni-dimensionnel qui est un syst\`{e}me exactement soluble. Un oscillateur en interaction avec une source externe est d\'{e}crit par le Lagrangien \be L = \frac{1}{2} m \dot{x}^2 - \frac{1}{2}m \omega^2 x^2 + J x.\ee La source d\'{e}pendante du temps $J(t)$ peut \^{e}tre vue, par exemple, comme un champ \'{e}lectrique si l'oscillateur est suppos\'{e} acqu\'{e}rir une charge \'{e}lectrique. La limite  de l'oscillateur harmonique libre s'obtient de ce syst\`{e}me par passage \`{a} la limite $J(t) \to 0$. L'\'{e}quation d'Euler-Lagrange appliqu\'{e}e \`{a} l'action~\eqref{action}donne la trajectoire classique : \be \label{extre} \frac{\delta S[x]}{\delta x(t)}=0\ee ce qui donne \be m \ddot{x}_{\mathrm{cl}}+ m \omega^2 x_{\mathrm{cl}}-J=0\ee

\begin{figure}[htbp]
\centering
  \includegraphics[width=5cm]{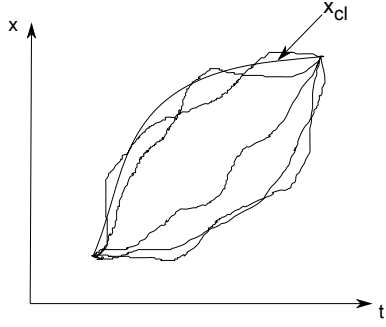}\\
  \caption{Fluctuations autour de la trajectoire classique.}
\end{figure}

La forme g\'{e}n\'{e}rale de l'amplitude de transition est, comme nous l'avons vu dans le chapitre 2,  est donn\'{e}e par \be U(t_f,x_f;t_f,t_i)=A \int \mathscr{D}[x(t)] \, e^{\frac{i}{\hbar}S[x(t)]} \ee Pour \'{e}valuer cette int\'{e}grale fonctionnelle, notons que l'action est quadratique dans les variables dynamiques $x(t)$. Par cons\'{e}quent, si on d\'{e}finit \be x(t)=x_{\mathrm{cl}}+\eta(t),\ee alors nous pouvons d\'{e}velopper l'action par rapport au chemin classique comme \begin{multline} \label{quadra}S[x(t)]=S[x_{\mathrm{cl}}+\eta(t)]=S[x_{\mathrm{cl}}]+\int \d t \, \eta(t) \frac{\delta S[x]}{\delta x(t)}\Big | _{x=x_{\mathrm{cl}}}\\ +\frac{1}{2!} \int \d t_1 \, \d t_2 \, \eta(t_1) \eta(t_2)\, \frac{\delta^2 S[x]}{\delta x(t_1) \delta x(t_2)}\Big |_{x=x_{\mathrm{cl}}}. \end{multline}Nous notons, d'apr\`{e}s~\eqref{extre}, que la trajectoire classique est un extremum pour l'action. Donc, nous avons \be \frac{\delta S[x]}{\delta x(t)} \Big |_{x=x_{\mathrm{cl}}}=0.\ee Par cons\'{e}quent, nous pouvons \'{e}crire l'\'{e}quation~\eqref{quadra} comme \be S[x]=S[x_{\mathrm{cl}}]+ \frac{1}{2!} \int \d t_1 \, \d t_2 \, \eta(t_1) \eta(t_2)\, \frac{\delta^2 S[x]}{\delta x(t_1) \delta x(t_2)}\Big |_{x=x_{\mathrm{cl}}}.\ee La d\'{e}riv\'{e}e fonctionnelle dans cette \'{e}quation peut \^{e}tre \'{e}valu\'{e}e pour l'action de l'oscillateur harmonique. Un calcul direct faisant appel aux propri\'{e}t\'{e}s des d\'{e}riv\'{e}es fonctionnelles (voir chapitre 1) donne
\be S[x]=S[x_{\mathrm{cl}}]+ \frac{1}{2} \int_{t_i}^{t_f} \d t \, m \left(\dot{\eta}^2- \omega^2 \eta^2\right)\ee

Les variables $\eta(t)$ repr\'{e}sentent les fluctuations quantiques autour du chemin classique, c'est \`{a} dire elles mesurent la d\'{e}viation d'une trajectoire par rapport \`{a} la trajectoire classique. Puisque  les extr\'{e}mit\'{e}s des trajectoires sont fix\'{e}s, les fluctuations satisfont aux conditions de bords :\be \label{bord}\eta(t_i)=\eta(t_f)=0\ee

Il est clair que la sommation sur tous les chemins est \'{e}quivalente \`{a} celle sur toutes les fluctuations  possibles sujettes \`{a} la condition~\eqref{bord}. Par cons\'{e}quent, \bea \label{oshar}\nonumber U(t_f,x_f;t_i,x_i)&=&A \int \mathscr{D}\eta \, e^{\frac{i}{\hbar}S[x_{\mathrm{cl}}]+ \frac{i}{2 \hbar}\int_{t_i}^{t_f}\d t \, m \left(\dot{\eta}^2- \omega^2 \eta^2\right)}\\
&=& A e^{\frac{i}{\hbar} S[x_{\mathrm{cl}}]}\int \mathscr{D}\eta \, e^{\frac{i}{2 \hbar}\int_{t_i}^{t_f}\d t \, m \left(\dot{\eta}^2- \omega^2 \eta^2\right)}\eea  Nous remarquons que l'exposant dans cette int\'{e}grale est quadratique par rapport aux variables d'int\'{e}gration. De telles int\'{e}grales peuvent \^{e}tre \'{e}valu\'{e}es de plusieurs fa\c{c}ons. Nous proposons ici une m\'{e}thode bas\'{e}e sur un d\'{e}veloppement de Fourier.

Dans l'int\'{e}grale~\eqref{oshar}, on red\'{e}finit la variable temporelle comme \[t \to t-t_i.\] Donc \be U(t_f,x_f;t_i,x_i) = A e^{\frac{i}{\hbar} S[x_{\mathrm{cl}}]}\int \mathscr{D} \eta \, e^{\frac{i}{2 \hbar}\int_{0}^{T}\d t \, \left(m \dot{\eta}^2- m \omega^2 \eta^2\right)}\ee o\`{u} $T=t_f - t_i$. La variable $\eta(t)$ satisfait \`{a} la condition aux limite \be \eta(0)=\eta(T)=0.\ee La fluctuation $\eta$ peut, par cons\'{e}quent, \^{e}tre d\'{e}velopp\'{e}e dans une base de fonctions trigonom\'{e}triques. Par exemple \be \eta(t)=\sum_n a_n \sin \left(\frac{n \pi t}{T}\right), \quad n=1,2, \ldots , N-1\ee $N$ \'{e}tant toujours le nombre de subdivisions de la trajectoire segment\'{e}e du chapitre pr\'{e}c\'{e}dent. Rempla\c{c}ant, on trouve \bea \nonumber \int_0^T \d t \, \dot{\eta}^2(t) &=&\sum_{n,m}\int_0^T \d t \, a_n a_m \frac{n m \pi^2}{T^2} \cos \left(\frac{n \pi t}{T}\right) \cos \left(\frac{m \pi t}{T}\right)\\
&=& \frac{T}{2}\sum_n \left(\frac{n \pi }{T}\right)^2 a_n^2 ,\eea o\`{u} nous avons utilis\'{e} la propri\'{e}t\'{e} d'orthogonalit\'{e} de la fonction cosinus. Nous trouvons de m\^{e}me,
\bea \nonumber \int_0^T \d t \, \eta^2(t) &=& \sum_{n,m}\int_0^T \d t \, a_n a_m \sin \left(\frac{n \pi t}{T}\right) \sin \left(\frac{m \pi t}{T}\right)\\
&=& \frac{T}{2}\sum_n a_n^2 .\eea Notons que l'int\'{e}grale sur les variables $\eta(t)$ est \'{e}quivalente \`{a} l'int\'{e}gration sur les variables $a_n$ qui sont les coefficients de Fourier . On peut par cons\'{e}quent \'{e}crire \begin{multline}\label{u}U(t_f,x_f;t_i,x_i)  \\ = \lim_{{\epsilon \to 0\atop N \to \infty}} A' e^{\frac{i}{\hbar} S[x_{\mathrm{cl}}]}\int_{} \d a_1 \cdots \d a_{N-1} \, e^{\frac{i}{2 \hbar}\sum_{n=1}^{N-1}\left(\frac{T}{2}\left(\frac{n \pi}{T}\right)^2 m a_n^2-\frac{T}{2}m \omega^2 a_n^2\right)}\\
= \lim_{{\epsilon \to 0\atop N \to \infty}} A' e^{\frac{i}{\hbar} S[x_{\mathrm{cl}}]}\int_{} \d a_1 \cdots \d a_{N-1} \, e^{\frac{i m T}{4 \hbar}\sum_{n=1}^{N-1}\left(\left(\frac{n \pi}{T}\right)^2 - \omega^2\right) a_n^2}
\end{multline} Ici, nous notons que le Jacobien du changement de variables de $\eta$ aux coefficients $a_n$ a \'{e}t\'{e} "absorb\'{e}" dans la constante $A'$ dont la forme explicite sera d\'{e}termin\'{e}e par la suite.

L'amplitude de transition est, dans ce cas,  un produit d'int\'{e}grales de Fresnel que l'on peut ais\'{e}ment \'{e}valuer. En fait, nous avons \bea \nonumber\int_{} \d a_n e^{i \frac{m T}{4 \hbar}\left(\left(\frac{n \pi}{T}\right)^2-\omega^2\right)a_n^2}&=&\left(\frac{4 \pi i \hbar}{m T}\right)^{1/2}\left(\left(\frac{n \pi}{T}\right)^2-\omega^2\right)^{-1/2}\\
&=&\left(\frac{4 \pi i \hbar}{m T}\right)^{1/2} \frac{T}{n \pi}\left(1-\left(\frac{\omega T}{n \pi}\right)^2\right)^{-1/2}.\eea Substituant cette forme dans l'expression~\eqref{u}, nous obtenons
\be U(t_f,x_f;t_i,x_i) = \lim_{{\epsilon \to 0\atop N \to \infty}} A'' e^{\frac{i}{\hbar} S[x_{\mathrm{cl}}]} \prod_{n=1}^{N-1}\left(1-\left(\frac{\omega T}{n \pi}\right)^2\right)^{-1/2}, \ee la constante $A''$ ``absorbe'' toutes les constantes. Si nous utilisons maintenant l'identit\'{e}, \be \prod_{n=1}^\infty \left(1-\frac{u^2}{n^2}\right)=\frac{\sin \pi u}{\pi u},\ee nous obtenons apr\`{e}s passage \`{a} la limite $N \to \infty$ \be \label{harmonique}U(t_f,x_f;t_i,x_i) = \lim_{{\epsilon \to 0\atop N \to \infty}} A''  e^{\frac{i}{\hbar} S[x_{\mathrm{cl}}]} \left(\frac{\sin \omega T}{\omega T}\right)^{-1/2}.\ee Nous pouvons calculer la constante $A''$ simplement par noter que lorsque $\omega \to 0$ , l'oscillateur harmonique se r\'{e}duit \`{a} une particule libre. En fait, identifiant l'\'{e}quation~\eqref{libre} et l'\'{e}quation~\eqref{harmonique} pour $\omega=0$, nous obtenons \be \lim_{{\epsilon \to 0\atop N \to \infty}} A'' = \left(\frac{m}{2 \pi i \hbar T}\right)^{\frac{1}{2}}. \ee La forme compl\`{e}te de l'amplitude de transition peut finalement s'\'{e}crire sous la forme
\be U(t_f,x_f;t_i,x_i) =\left(\frac{m \omega}{2\pi i \hbar \sin \omega T}\right)^{\frac{1}{2}}e^{\frac{i}{\hbar} S[x_{\mathrm{cl}}]}.\ee Quant \`{a} l'action classique $S[x_{\mathrm{cl}}]$ de l'oscillateur harmonique, elle peut \^{e}tre directement d\'{e}duite du Lagrangien : \be S[x_{\mathrm{cl}}]=\frac{m \omega}{2 \sin \omega T}\left\{(x_i^2+x_f^2) \cos \omega T - 2x_i x_f\right\}\ee

\chapter{ M\'{e}thode de l'int\'{e}grale de chemin pour la th\'{e}orie quantique des champs}
\section{Introduction} Nous allons voir dans ce chapitre l'application la plus int\'{e}ressante des int\'{e}grales de chemin. Comme nous l'avons mentionn\'{e} dans l'introduction, cette m\'{e}thode est surtout bien adapt\'{e}e \`{a} l'\'{e}tude de syst\`{e}mes \`{a} tr\`{e}s grand nombre de degr\'{e}s de libert\'{e}, voire une infinit\'{e} comme pour la th\'{e}orie des champs.

Notre objectif est de pr\'{e}senter une explication \'{e}l\'{e}mentaire des math\'{e}matiques derri\`{e}re les int\'{e}grales de chemin de Feynman appliqu\'{e}es \`{a} la th\'{e}orie quantique des champs (TQC). Notre prototype est une TQC connue sous le nom de th\'{e}orie des champs scalaires avec terme d'interaction en $\phi^4$. Quoique l'exemple est \'{e}l\'{e}mentaire, il fournit suffisamment d'informations pour comprendre l'application des int\'{e}grales de chemin aux syst\`{e}mes de plusieurs particules en interaction.

Le passage de la m\'{e}canique quantique \`{a} la TQC est direct, mais le concept sous-jacent est un peu compliqu\'{e}e \`{a} saisir. Fondamentalement, trois questions se posent \`{a} ce niveau : (1). En TQC, on exige que la dynamique de la th\'{e}orie soit relativiste ; (2). Les variables de l'espace et du temps doivent \^{e}tre trait\'{e}es aux m\^{e}mes pieds d'\'{e}galit\'{e}, ce qui est pr\'{e}cis\'{e}ment une autre exigence de la th\'{e}orie de la relativit\'{e}. En m\'{e}canique quantique, le temps est juste un param\`{e}tre alors que la position est un op\'{e}rateur (le symbole $\ket{t}$ est d\'{e}nu\'{e} de sens); (3). La m\'{e}canique quantique est avant tout une th\'{e}orie \`{a} une particule, alors que dans la TQC plusieurs particules se mettent en jeu. L'int\'{e}grale de chemin r\'{e}pond admirablement \`{a} toutes ces exigences.

Maintenant, voici le grand saut en un mot : la TQC remplace la coordonn\'{e}e de position $x$ par un champ $\phi(x)$, o\`{u} la quantit\'{e} $x$ est actuellement un raccourci pour les coordonn\'{e}es $x, y, z, t$. Ce processus de r\'{e}affectation des coordonn\'{e}es est connu sous le nom de \emph{seconde quantification}.

En TQC, nous supposons l'existence d'un champ quantique $\phi(x)$ qui pourra d\'{e}crire des sp\'{e}cificit\'{e}s comme le spin des particules, moment angulaire, etc.  Dans ce qu'on appelle TQC canonique, le champ lui m\^{e}me est un op\'{e}rateur (heureusement, les int\'{e}grales de chemin ignorent cette complication). A partir de ce que nous avons vu dans le chapitre pr\'{e}c\'{e}dent, il est tout \`{a} fait compr\'{e}hensible que l'int\'{e}grale de chemin dans la TQC soit de la forme :
\be Z= \bigintsss_{-\infty}^{+\infty}\mathscr{D} \phi \, \exp{\left\{\frac{i}{\hbar} \int_{-\infty}^{+\infty}L(\phi , x^\mu , \partial_\mu \phi) \d ^4 x\right\}}\ee o\`{u} tous les coefficients libres sont actuellement ``absorb\'{e}s'' dans la mesure $\mathscr{D} \phi$ qui prend une forme comme $\d \phi(x_1)\, \d \phi(x_2) \cdots $. Le fait d'appeler cette int\'{e}grale $Z$ est une pure convention. D'une fa\c{c}on imag\'{e}e, on pourrait voir cette int\'{e}grale comme l'amplitude de transition pour un champ qui se propage du ``vide'' \`{a} $t=-\infty$ au ``vide'' \`{a} $t=\infty$. Un champ peut g\'{e}n\'{e}ralement d\'{e}crire n'importe quelle dynamique, mais nous pouvons le voir dans notre situation comme une quantit\'{e} d\'{e}crivant une population de particules s'interagissant en chaque point de l'espace-temps. Il faut noter qu'appeler $Z$ une int\'{e}grale de chemin n'est pas appropri\'{e} puisqu'elle ne s'\'{e}tend pas sur les chemins de l'espace-temps. On l'appelle l'int\'{e}grale fonctionnelle $Z$ ou la fonctionnelle g\'{e}n\'{e}ratrice $Z$ (generating functional).

Parce que l'int\'{e}grale fonctionnelle avec termes d'interaction ne peut pas \^{e}tre \'{e}valu\'{e}e directement, on a recours en g\'{e}n\'{e}ral \`{a} une approche perturbative.

La forme du Lagrangien pour un champ d\'{e}pend de la nature des particules et des forces qui interviennent dans l'interaction. Par cons\'{e}quent, il y a des Lagrangiens pour des particules scalaires (spin z\'{e}ro) ou bosons, pour les spineurs (spin {\small 1/2}) ou fermions et pour des particules vectorielles (spin 1). Il y a m\^{e}me un pour les gravitons (spin 2). Le plus simple est le Lagrangien scalaire ou bosonique. Le pr\'{e}sent chapitre s'illustrera sur une TQC scalaire relativiste ayant pour Lagrangien : \be L={1 \over 2} \left(\partial_\mu \phi \partial^\mu \phi - m^2 \phi^2\right)-V\ee Le terme d'interaction le plus simple que l'on peut consid\'{e}rer est $V \sim \lambda  \phi^4$ o\`{u} $\lambda$ est une constante appel\'{e}e la constante de couplage. C'est ce qu'on appelle une TQC \`{a} self-interaction (self-interacting field theory), autrement dit, le champ interagit avec lui-m\^{e}me et avec toute particule cr\'{e}\'{e}e le long de l'interaction. Comme nous l'avons indiqu\'{e}, l'int\'{e}grale $Z$ n'a pas une forme explicite et de ce fait un traitement perturbatif s'impose. Ceci conduit \`{a} une interpr\'{e}tation fort int\'{e}ressante du processus de cr\'{e}ation et de propagation des particules.

Le probl\`{e}me est donc de r\'{e}soudre l'int\'{e}grale \be Z(\lambda)=\bigintsss_{} \mathscr{D} \phi \exp \left\{\frac{i}{\hbar}\int_{} \left[{1 \over 2} \left(\partial_\mu \phi \partial^\mu \phi - m^2 \phi^2 \right)- \lambda \phi^4 \right] d^4 x \right\}\ee cette int\'{e}grale ne peut pas \^{e}tre attaqu\'{e}e dans cette forme. Le probl\`{e}me principal est que le nombre de variables d'int\'{e}gration est infini. Nous aurons besoin de faire quelques changements avant de proposer une solution perturbative.

\section{Modification de la fonctionnelle g\'{e}n\'{e}ratrice $Z$} Consid\'{e}rons la forme libre ($\lambda=0$) de $Z$ avec une ``source'' $J(x)$ :  \be \label{0} Z[J]=\bigintsss_{} \mathscr{D} \phi \, \exp \left\{i \int_{} \, {1 \over 2} \left(\partial_\mu \phi \partial^\mu \phi - m^2 \phi^2 \right) + J(x) \phi  \; \d x \right\}. \ee $J$ tendra vers 0 par la suite. Notons que nous nous pla\c{c}ons dans un syst\`{e}me d'unit\'{e} o\`{u} $\hbar = c=1$ et que $\d x$ remplace $\d^4 x$ pour all\'{e}ger les \'{e}critures.  Une int\'{e}gration par partie du terme $\partial_\mu \phi \partial^\mu \phi$ donne \[\int  {1 \over 2} \left(\partial_\mu \phi \partial^\mu \phi - m^2 \phi^2 \right) \, \d x=-\int \left( {1 \over 2} \phi \partial^2 \phi + {1 \over 2} m^2 \phi^2 \right) \, \d x\]
Maintenant, supposons que le champ peut \^{e}tre mis sous la forme \[\phi(x)=\phi_0(x)+\varphi(x)\] o\`{u} $\phi_0(x)$ est ce qu'on appelle la solution ``classique'' de l'\'{e}quation h\'{e}t\'{e}rog\`{e}ne \be -\left(\partial^2+m^2\right)\phi_0(x)=J\ee c'est l'\'{e}quation de Klein-Gordon avec source. La solution classique est unique (elle correspond au ``chemin'' classique). Elle peut \^{e}tre obtenue par la m\'{e}thode usuelle des fonctions de Green :
\be \label{green} -\left(\partial^2+m^2\right)G(x-x')=\delta^4 (x-x')\ee o\`{u} $G(x-x')$ est la fonction de Green (\`{a} 4 dimensions) associ\'{e}e \`{a} l'op\'{e}rateur $-\left(\partial^2+m^2\right)$. Donc \[ \phi_0(x) = \int G(x-x') J(x') \d^4 x'\] est la solution d\'{e}sir\'{e}e. Pour r\'{e}soudre~\eqref{green} par rapport \`{a} $G(x-x')$, nous supposons que la fonction de Green peut \^{e}tre exprim\'{e}e comme une transform\'{e}e de Fourier d'une certaine fonction $G(k)$ : \be G(x-x')=\int \frac{\d^4 k}{(2 \pi)^4}G(k) e^{i k \cdot (x-x')}\ee o\`{u} $k$ est le quadrivecteur \'{e}nergie-impulsion $(E/c, \overrightarrow{p})$. En faisant agir l'op\'{e}rateur $-\left(\partial^2+m^2\right)$ sur l'expression de $G(x-x')$, et tenant compte du fait que la transform\'{e}e de l'unit\'{e} est tout simplement la distribution de Dirac : \[\delta^4(x-x')=\int \frac{\d^4 k}{(2 \pi)^4} e^{i k \cdot (x-x')}, \] nous aboutissons \`{a} la solution \be \label{prop} G(k) = \frac{1}{k^2-m^2},
 \ee et par suite, \bea \label{green2}G(x-x') &=&  \int_{} \frac{\d^4 k}{(2 \pi)^4}\frac{e^{i k \cdot (x-x')}}{k^2-m^2}\\
 \phi_0(x) &= &\int G(x-x') J(x') \d^4 x' .\eea Dans la TQC scalaire, la fonction de Green~\eqref{green2} est appel\'{e}e le propagateur de Feynman et not\'{e}e $\Delta_F (x-x')$. Le propagateur $G(k)$ exprim\'{e} dans l'espace des impulsions nous sera utile plus tard quand nous d\'{e}finirons les r\`{e}gles dites de Feynman pour un processus de diffusion~\footnote{Pour \'{e}viter la divergence de l'int\'{e}grale~\eqref{green2}, le d\'{e}nominateur de l'int\'{e}grand est g\'{e}n\'{e}ralement remplac\'{e} par $k^2-m^2+i \epsilon$.}. Pour r\'{e}capituler, nous avons les formules: \[\boxed{  \phi_0(x) = \int \Delta_F (x-x') J(x') \d^4 x' ,}\] o\`{u}
 \[ \boxed{ \Delta_F(x-x')=\int \frac{\d^4 k}{(2 \pi)^4}\frac{e^{i k \cdot (x-x')}}{k^2-m^2}.}\] Un calcul direct mais long permet d'\'{e}crire~\eqref{0} sous la forme \[ Z[J]=\int_{} \mathscr{D} \varphi \, \exp \left\{\frac{i }{2}\left( \int_{} \, \left(\partial_\mu \varphi \partial^\mu \varphi - m^2 \varphi^2 \right)\d^4 x - \iint J(x')\Delta_F(x''-x') J(x'') \d^4 x' \d^4 x'' \right)\right\}.\] Faisant sortir tout ce qui ne d\'{e}pend pas de $\varphi$, on a \[Z[J]=N  \exp \left\{-\frac{i }{2} \iint J(x')\Delta_F(x''-x') J(x'') \d^4 x' \d^4 x'' \right\}.\]
o\`{u}
\[N=\int_{} \mathscr{D} \varphi \, \exp \left\{\frac{i }{2}\int_{} \, \left(\partial_\mu \varphi \partial^\mu \varphi - m^2 \varphi^2 \right)\d^4 x\right\}\]  peut \^{e}tre vue comme une constante de normalisation, mais nous allons poser $N = 1$ parce que nous utiliserons des amplitudes normalis\'{e}es plus tard. On a donc finalement
\be \label{ZJ}\boxed{Z[J]=\exp \left\{-\frac{i }{2} \iint J(x)\Delta_F(x-x') J(x') \d^4 x \d^4 x' \right\}.}\ee
Nous avons donc transform\'{e} une int\'{e}grale \`{a} nombre infini de variables \`{a} une autre sur les quatre variables spatio-temporelles. Nous sommes maintenant en mesure de d\'{e}velopper la fonctionnelle g\'{e}n\'{e}ratrice $Z[J]$ pour un traitement perturbatif.

\section{D\'{e}veloppement en s\'{e}rie de puissances - les fonctions de Green}
La fonctionnelle $Z$ peut se d\'{e}velopper en s\'{e}rie de Taylor comme suit (conform\'{e}ment \`{a} la formule~\eqref{taylor})
\be Z[J]=\sum_{n=0}^\infty \int \frac{i^n}{n!} \; \d x_1 \d x_2 \cdots \d x_n J(x_1) J(x_2) \cdots J(x_n) G(x_1, \ldots , x_n)\ee o\`{u} \be \label{G} G(x_1, \ldots , x_n)= \frac{1}{i^n} \left[ \frac{\delta}{\delta J(x_1)} \frac{\delta}{\delta J(x_2)} \cdots \frac{\delta}{\delta J(x_n)}\right] Z[J] \Big|_{J=0}.\ee La fonction $G$ est une sorte de fonction de Green. Dans la TQC elle est appel\'{e}e \emph{fonction \`{a} $n$-points} ($n$-point function). Nous allons voir que la fonction \`{a} $n$-points n'est non-nulle que pour des valeurs paires de $n$. Les physiciens ont compris que ces fonctions ne sont autres que les amplitudes de transition. Par exemple, d\'{e}rivant~\eqref{ZJ}, nous pouvons calculer la fonction \`{a} 2 points \bea \nonumber G(x_1,x_2)&=&\frac{1}{i^2} \left[ \frac{\delta}{\delta J(x_1)} \frac{\delta}{\delta J(x_2)}\right] Z(J)\Big|_{J=0}\\
\nonumber &=& i \Delta_F (x_1-x_2).\eea Sch\'{e}matiquement, $\Delta_F (x_1-x_2)$ peut se repr\'{e}senter par un graphe connexe \`{a} 2 points.
\[\includegraphics[width=5cm]{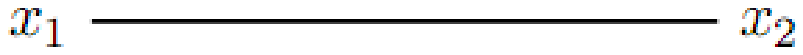}\]
Un calcul similaire donne pour la fonction \`{a} 4 points \[G(x_1,x_2,x_3,x_4)=-\Delta_F(x_1-x_2)\Delta_F(x_3-x_4)-\Delta_F(x_1-x_3)\Delta_F(x_2-x_4)-\Delta_F(x_1-x_4)\Delta_F(x_2-x_3)\] ce qui peut se repr\'{e}senter par les trois graphes \`{a} 4 points
\[\includegraphics[width=10cm]{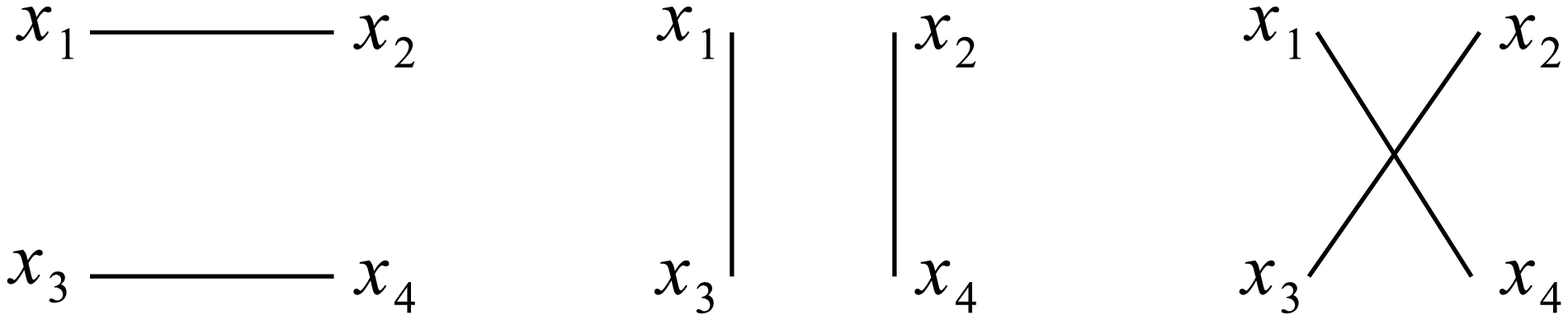} \] \vskip-30pt
Pour la fonction \`{a} 6 points, on trouve 15 termes correspondant \`{a} 15 graphes diff\'{e}rents. En g\'{e}n\'{e}ral, pour la fonction \`{a} $2n$ points, le nombre de graphes est le nombre de fa\c{c}ons de grouper les $2n$ points en paires distinctes. Une analyse combinatoire inductive montre que ce nombre est \'{e}gale \`{a} \be \label{cn}(2n-1)!! = 1\times 3 \times 5 \cdots \times (2n-1) = \frac{(2n-1)!}{2^{n-1}(n-1)!}=\frac{(2n)!}{2^{n}n!}.\ee Ce r\'{e}sultat peut se d\'{e}montrer par r\'{e}currence.

La fonctionnelle $Z$ est un exponentiel. Le calcul peut donc \^{e}tre fait pour n'importe quel ordre. On verra qu'en g\'{e}n\'{e}ral, le calcul est souvent laborieux. Heureusement, on verra aussi qu'il y a une formule simple pour ce genre de calcul perturbatif.

On interpr\`{e}te les fonctions $G$ comme des quantit\'{e}s associ\'{e}es \`{a} des particules cr\'{e}es \`{a} un certain point de l'espace-temp $x_i$, se propage le long d'une ligne (repr\'{e}sent\'{e}e par la quantit\'{e} $\Delta_F(x_f-x_i)$), et annihil\'{e}e ensuite en un point final $x_f$. Le propagateur de Feynman est le pivot des fonctions \`{a} $n$-points et de ce qu'on appelle les diagrammes de Feynman.

Feynman a compris que les multiples differentiations~\eqref{G} produit des termes que l'on peut d\'{e}crire par des graphes simples. Chaque terme a un coefficient qui lui est associ\'{e} relatif \`{a} la topologie du diagramme correspondant. L'ordre de multiplicit\'{e} de chaque terme d\'{e}pend du nombre de fa\c{c}ons avec lesquelles on peut tracer le diagramme. Ces amplitudes et multiplicit\'{e}s peuvent \^{e}tre \'{e}valu\'{e}s par l'analyse combinatoire.

\section{L'int\'{e}grale fonctionnelle $Z$ avec le terme d'interaction}
Nous allons maintenant introduire dans le Lagrangien le terme d'interaction $\lambda$ du potentiel $\phi^4$. Par une convention qui sera claire par la suite, nous poserons $V=\lambda \phi^4 /4!$. La nouvelle fonctionnelle $Z$ devient alors \bea  Z(\lambda)&=&\bigintsss_{} \mathscr{D} \phi \, \exp \left\{i  \int_{}  \, \left[\frac{1}{2}\left(\partial_\mu \phi(x) \partial^\mu \phi(x) - m^2 \phi^2(x)\right) -\frac{\lambda}{4!}\phi^4(x)  \right]\, \d^4 x \right\}\\
\nonumber &=& \bigintsss_{} \mathscr{D} \phi \, \exp \left\{i  \int_{}  \,\frac{1}{2}\left(\partial_\mu \phi(x) \partial^\mu \phi(x) - m^2 \phi^2(x)\right)\, \d^4 x  -i \frac{\lambda}{4!} \int_{} \phi^4(x)  \, \d^4 x\right\} \eea
Cette int\'{e}grale se pr\'{e}sente vaguement sous la forme \[\int e^{-u^2-bu^4}\, \d u.\] Il est facile de montrer que l'on peut \'{e}crire cette int\'{e}grale sous la forme diff\'{e}rentielle suivante :
\[\int e^{-u^2-bu^4}\, \d u = \exp\left[-b \frac{d^4}{d^4 c}\right] \int e^{-u^2+c u}\, \d u \Big|_{c=0}\] Nous rappelons que pour un op\'{e}rateur $O$,  $\exp (O) = \sum_{n=0}^\infty O^n/n! $. Nous allons user de cette approche pour introduire le terme exponentiel contenant $\lambda$ dans la fonctionnelle $Z[J]$. Nous \'{e}crivons $Z(\lambda)$ comme \be Z(\lambda)=\exp\left[-i \frac{\lambda}{4!} \int \d^4 x \frac{\delta^4}{i^4 \delta J(x)^4}\right] Z[J] \Big|_{J=0}.\ee Rempla\c{c}ant $Z[J]$ par son expression~\eqref{ZJ}, nous obtenons
\be \boxed{Z(\lambda)=\exp\left[-i \frac{\lambda}{4!} \int \d^4 x \frac{\delta^4}{i^4 \delta J(x)^4}\right]\exp \left\{-\frac{i }{2} \iint J(x)\Delta_F(x-x') J(x') \d^4 x \, \d^4 x' \right\}\Big|_{J=0}.}\ee L'exponentiel de l'op\'{e}rateur diff\'{e}rentiel se d\'{e}veloppe comme suit \begin{multline} \label{exp}\exp\left[-i \frac{\lambda}{4!} \int \d^4 x \frac{\delta^4}{ \delta J(x)^4}\right]= 1-i \frac{\lambda}{4!} \int \d^4 x \frac{\delta^4}{ \delta J(x)^4}-\frac{\lambda^2}{2! (4!)^2}\int \d^4 x \frac{\delta^4}{\delta J(x)^4} \int \d^4 y \frac{\delta^4}{ \delta J(y)^4}+ \ldots \\ = \sum_{n=0}^\infty \left(\frac{-i \lambda}{4!}\right)^n \frac{1}{n!}\int \d x_1 \frac{\delta^4}{ \delta J(x_1)^4} \times \int \d x_2 \frac{\delta^4}{ \delta J(x_2)^4} \times \cdots \int \d x_n \frac{\delta^4}{\delta J(x_n)^4}.\end{multline}

\'{E}tudions en d\'{e}tail les d\'{e}riv\'{e}es fonctionnelles de $Z[J]$ par rapport \`{a} $J$. On a
\begin{multline} \frac{\delta}{ \delta J(x_1)}\left[ -\frac{i}{2}\int J(x') \Delta_F (x'-x'') J(x'') \d x' \d x''\right]\\= -\frac{i}{2} \int \delta(x'-x_1)\Delta_F (x'-x'')J(x'') \d x' \d x'' - \frac{i}{2}\int J(x') \Delta_F (x'-x'') \delta(x''-x_1) \d x' \d x''\\
= -i \int \Delta_F (x_1-x') J(x') \d x'
\end{multline}
Donc \[\frac{\delta Z}{\delta J(x_1)}=\left(-i \int \Delta_F (x_1-x') J(x') \d x'\right) Z.\] Nous avons de m\^{e}me \[\frac{\delta^2 Z}{\delta J(x_1) \delta J(x_2)} = -i \Delta_F (x_1-x_2)Z-\left(i \int \Delta_F (x_1-x') J(x') \d x'\right) \frac{\delta Z}{\delta J(x_2)}.\] Il est commode d'adopter la notation concise suivante : $$Z_1\equiv\delta Z/ \delta J(x_1), \quad  Z_{12}\equiv \delta^2 Z/ \delta J(x_1) \delta J(x_2), \hbox{etc} \quad \Delta_{12}\equiv\Delta_F(x_1-x_2). $$ En \'{e}liminant l'int\'{e}grale, nous pouvons \'{e}crire \[Z_{12}=-i \Delta_{12} Z + \frac{1}{Z}Z_1 Z_2. \]
Seuls les termes qui sont proportionnels \`{a} $Z$ survivent dans la limite $J \to 0$. Il est aussi important de souligner que tous les termes avec nombres de diff\'{e}renciations impair s'annulent dans cette limite (comme $Z_{12345}$).

Pour \'{e}valuer la fonction \`{a} $n$ points $G(x_1, \ldots , x_n)$, nous devons \'{e}valuer des d\'{e}rivations de $Z(\lambda)$ aux points $x_1, x_2, \ldots , x_n$. Par exemple, la fonction \`{a} 4 points au premier ordre par rapport \`{a} $\lambda$ dans le d\'{e}veloppement~\eqref{exp} sera $$Z_{xxxx1234}\big|_{J=0} = \frac{\delta^6 Z}{\delta J(x)^4 \delta J(x_1) \delta J(x_2) \delta J(x_3)\delta J(x_4)}\Big|_{J=0}$$ o\`{u} $x$ est le point d'interaction. La fonction \`{a} 4 points \`{a} l'ordre 2 est donn\'{e}e par $Z _{xxxxyyyy1234}\big|_{J=0}$. L\`{a}, nous avons deux points d'interaction $x$ et $y$, etc.

\section{Formules de diff\'{e}rentiations et facteurs de sym\'{e}trie} Rappelons que la fonction \`{a} $n$ points $G(x_1, \ldots , x_n)$ se calcule par d\'{e}rivation fonctionnelle de $Z(\lambda)$ $n$ fois, \'{e}valu\'{e}es pour $J=0$. Pour la fonction \`{a} $2n$ points, nous avons d\'{e}j\`{a} vu qu'elle se pr\'{e}sente sous la forme g\'{e}n\'{e}rale \be \label{2n} Z_{123\ldots 2n}\Big|_{J=0}=(-i)^n \left(\Delta_{12} \Delta_{34} \cdots \Delta_{2n-1\;2n} + (C_n-1) \, \hbox{autres termes}\right)\ee o\`{u} le nombre $C_n$ est donn\'{e} par~\eqref{cn}. Comme nous l'avons expliqu\'{e}, chaque terme peut \^{e}tre repr\'{e}sent\'{e} par un graphe connexe. Ainsi, tout terme avec un nombre impair de d\'{e}riv\'{e}es fonctionnelles doit s'annuler vu que toute ligne dans le graphe doit joindre seulement deux points. La situation devient plus int\'{e}ressante lorsque l'on doit consid\'{e}rer des propagateurs de la forme $\Delta_F(x-x)=\Delta_F(0)$, o\`{u} les points initial et final sont les m\^{e}mes. Par exemple, calculons $Z_{xxxx}$ correspondant \`{a} $Z_{1234}$ lorsque $x_1=x_2=x_3=x_4=x$. Dans ce cas, nous avons \be \label{2vac} Z_{xxxx}\Big|_{J=0}=-3 \Delta_F(0)^2\ee Ryder~\cite{ryder} exprime graphiquement $\Delta_F(0)$ par un cercle \,\setlength{\unitlength}{1mm}
\begin{picture}(0, 0)
\put(0,1){\circle{4}}
\end{picture}\; \,pour d\'{e}signer qu'une particule se cr\'{e}e au point $x$, se propage pendant un moment et ensuite se d\'{e}truit au m\^{e}me point $x$, ce qui peut se pr\'{e}senter par une boucle; l'analogie est tr\`{e}s suggestive. Le terme~\eqref{2vac} se sch\'{e}matise par \, -3\;\setlength{\unitlength}{1mm}
\begin{picture}(0, 0)
\put(0,1){\circle{4}}
\put(4,1){\circle{4}}
\end{picture} \; \; \;. Le facteur de pond\'{e}ration -3 est appel\'{e} le facteur de sym\'{e}trie pour $Z_{xxxx}$. Sa valeur absolue repr\'{e}sente le nombre de fa\c{c}ons avec lesquelles un graphe peut \^{e}tre \'{e}crit. En g\'{e}n\'{e}ral, quand le nombre de d\'{e}rivations est assez grand, le calcul devient excessivement laborieux. Y-a-t-il un moyen d'\'{e}crire directement toutes ces quantit\'{e}s? La r\'{e}ponse est oui et c'est en plus relativement simple.

De l'\'{e}quation~\eqref{2n}, nous voyons qu'en m\^{e}me point $x$, nous avons $$Z_{\underbrace{xx \ldots x}_{2n \; \hbox{\tiny fois}}}=(-i)^n C_n \Delta_F(0)^n.$$ Le facteur de sym\'{e}trie dans ce cas est donn\'{e} par $(-i)^n C_n$. Mais qu'en-est-il lorsqu'on d\'{e}rive en plus par rapport \`{a} des variables $x_i$ qui ne sont pas n\'{e}cessairement \'{e}gales, par exemple \[Z_{xx12}\Big|_{J=0}=-\Delta_{F}(0)\Delta_{12}-2\Delta_{x1}\Delta_{x2}\] (o\`{u} $\Delta_{x1}=\Delta_F(x-x_1)$, etc). En g\'{e}n\'{e}ral, dans la quantit\'{e} \be \label{2n2p} Z_{\underbrace{xx \ldots x}_{2n \hbox{ \hbox{\tiny fois}}}12 \ldots 2p}\Big|_{J=0},\ee il y a, entre autres, des termes qui contiennent des facteurs $\Delta_F (0)^k$ pour $k \leq p$; les facteurs de sym\'{e}trie correspondant \`{a} ces termes sont donn\'{e}s par \be \label{fsym} (-i)^{p+n} \frac{(2n)!}{2^k k!}.\ee En fait, l'entier $k$ est \'{e}gale au nombre de boucles dans le graphe correspondant. Pour illustrer, un calcul de $Z_{xxxx1234}\big|_{J=0}$, qui, rappelons le, donne la fonction \`{a} 4 points \'{e}valu\'{e}e \`{a} l'ordre 1 par rapport \`{a} $\lambda$, donne \begin{multline} \label{coq} Z_{xxxx1234}\Big|_{J=0}=3 \Delta_F(0)^2 (\Delta_{12}\Delta_{34}+\Delta_{23}\Delta_{14}+\Delta_{13}\Delta_{24})\\ +12 \Delta_F(0)(\Delta_{x1}\Delta_{x2}\Delta_{34}+\Delta_{x2}\Delta_{x4}\Delta_{13}+ \Delta_{x3}
\Delta_{x4}\Delta_{12}\\ +\Delta_{x2}\Delta_{x3}\Delta_{14}+ \Delta_{x1}\Delta_{x4}\Delta_{23}
+\Delta_{x1}\Delta_{x3}\Delta_{24}) +24 \Delta_{x1}\Delta_{x2}\Delta_{x3}\Delta_{x4}.\end{multline}Nous remarquons que les termes proportionnels \`{a} $\Delta_F(0)^2$, $\Delta_F(0)$ et $\Delta_F(0)^0$ sont bien pond\'{e}r\'{e}s par~\eqref{fsym} pour $k=2,1$ et 0 respectivement, et de surcro\^{\i}t apparaissent avec des multiplicit\'{e}s respectivement \'{e}gales \`{a} 3, 6 et 1. Ces multiplicit\'{e}s s'expliquent par le fait que les points $x_i$ peuvent \^{e}tre permut\'{e}s pour produire des graphes \'{e}quivalents. Les trois termes de la quantit\'{e}~\eqref{coq} se sch\'{e}matisent par les diagrammes suivants \[  \includegraphics[width=15cm]{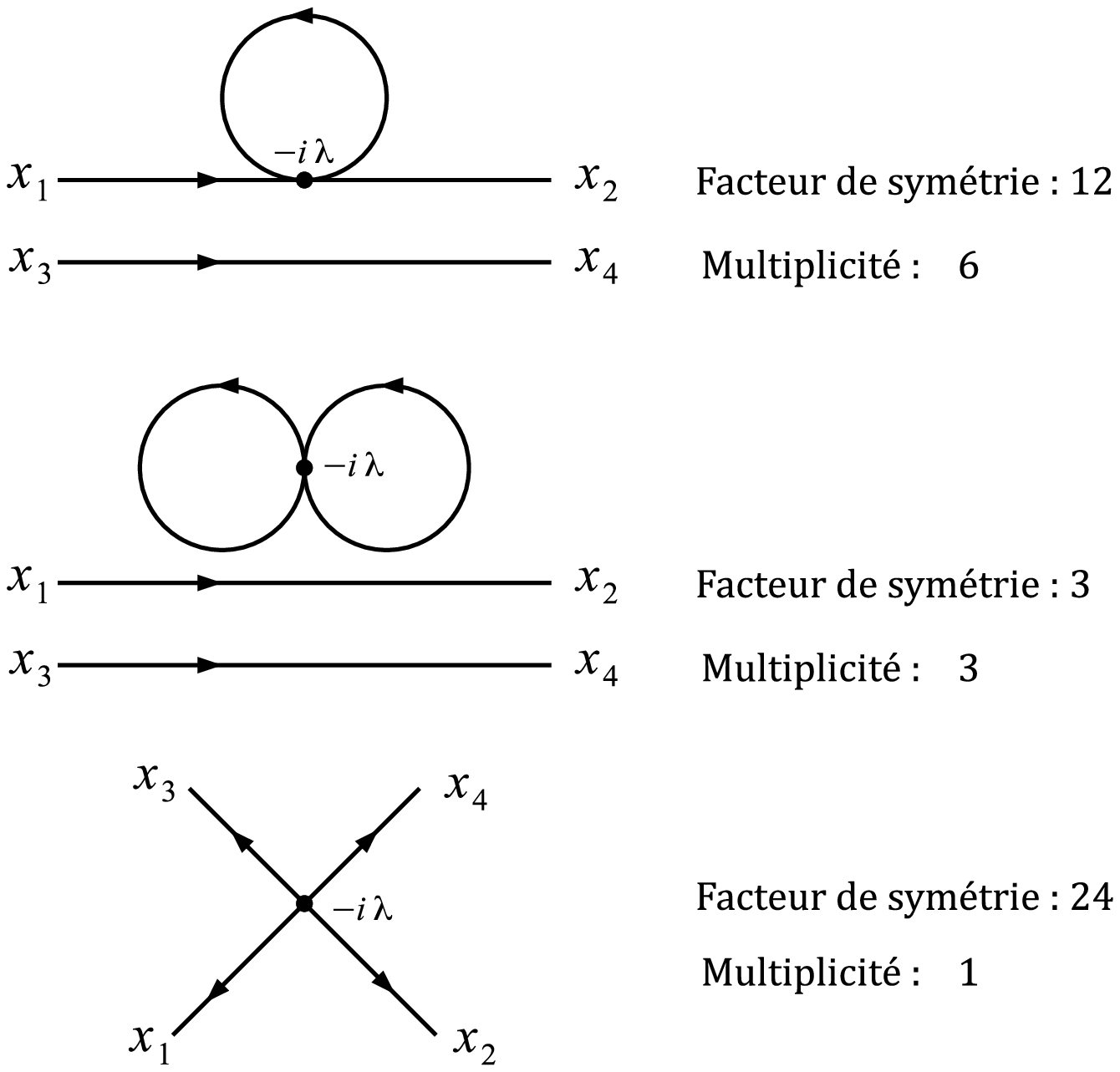}\]

Une analyse combinatoire inductive montre que la multiplicit\'{e} $M$ pour tout terme dans~\eqref{2n2p} correspondant \`{a} un nombre de boucles $k$, s'\'{e}crit
\be M=\frac{(2p)!}{2^t t! (2p-2t)!}\ee o\`{u} \be t=k+p-n.\ee Il ne faut pas oublier que c'est une formule valable uniquement \`{a} l'ordre $O(\lambda)$.

L'ordre $n$ par rapport \`{a} $\lambda$ fait, en g\'{e}n\'{e}ral, intervenir $n$ d\'{e}riv\'{e}es quadruples par rapport \`{a} $x_1, x_2, \ldots  x_n$. Ainsi, \`{a} l'ordre 2, on a \`{a} calculer $Z_{xxxxyyyy12 \ldots 2p}\big|_{J=0}$. Un exemple de graphes \`{a} 2 points, \`{a} l'ordre 2 est \[\includegraphics[width=10cm]{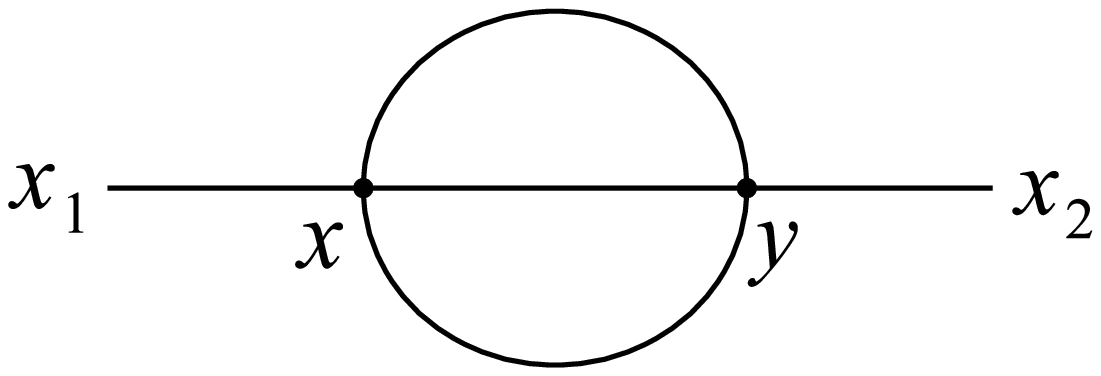} \vspace{-0.8cm}\] \vskip-0.5cm La formule donnant le facteur de sym\'{e}trie et la multiplicit\'{e} pour un ordre $n$ quelconque ($n$ vertexes internes)  et $2p$ points externes, s'\'{e}crit
\bea \label{cng}\nonumber C&=& \prod_{i,j=1 (i \leq j)}^n \frac{1}{2^{s_{ii}}s_{ij}!}\\
\nonumber M&=& \frac{(2p)!}{2^t t! (2p-2t)!}\\
t &=& p-2n+\sum_{i,j=1 (i \leq j)}^n s_{ij}.\eea $s_{ij}$ d\'{e}signe le nombre de lignes \emph{internes} joignant les vertexes $i$ et $j$. Par exemple, pour le diagramme pr\'{e}c\`{e}dent, on a $n=2$, $p=1$, $s_{xx}=s_{yy}=0$, $s_{xy}=3$, $t=0$ et $M=1$. Donc \[C=\frac{1}{3!}=\frac{1}{6}\]

\section{R\`{e}gles de Feynman}
Pour calculer l'amplitude de transition pour un processus r\'{e}el, Feynman a \'{e}labor\'{e} un ensemble de r\`{e}gles qui portent maintenant son nom. Rappelons~\eqref{prop}, qui est le propagateur de Feynman dans l'espace des moments $k$. C'est effectivement beaucoup plus pratique que d'utiliser la forme spatiale $\Delta_{xy}$, et nous pouvons utiliser~\eqref{prop} pour associer \`{a} chaque ligne dans un diagramme de Feynman avec un quadri-vecteur impulsion $k$. Les r\`{e}gles de Feynman pour la TQC $\phi^4$ sont
\begin{enumerate}
  \item Dessiner tous les diagrammes possibles correspondant au nombre d\'{e}sir\'{e} d'interactions ($x$, $y$, \ldots) et de points de l'espace-temps ($x_1$, $x_2$, \ldots), utilisant le temps comme l'axe vertical et l'espace comme l'axe horizontal. Pour chaque graphique, \'{e}tiqueter chaque ligne interne et externe avec une impulsion $k$ et l'orienter par une fl\`{e}che (le sens de la fl\`{e}che peut \^{e}tre compl\`{e}tement arbitraire). Pour chaque ligne interne, \'{e}crire l' int\'{e}grale propagateur \[\int \frac{\d k}{(2 \pi)^4}\frac{i}{k^2-m^2}.\]
  \item Pour chaque vertex d'interaction, \'{e}crire un facteur $-i \lambda$.
  \item Pour chaque vertex d'interaction, \'{e}crire une fonction delta de Dirac qui exprime la conservation d'impulsion sur ce sommet : la somme des 4-impulsions entrantes est \'{e}gale \`{a} la somme des 4-impulsions sortantes :\[(2\pi)^4 \delta\left(\sum_i k_i\right),\] o\`{u} $k$ est positif si la ligne entre le vertex et n\'{e}gative si elle sort du sommet.
  \item Pour chaque graphique, il y aura une fonction delta r\'{e}siduelle de la forme $(2\pi)^4 \delta(k_1 + k_2 + \ldots)$ qui exprime la conservation globale de la 4-impulsion dans le diagramme. \'{E}liminer ce terme.
  \item Dans l'\'{e}valuation des int\'{e}grales qui restent les fonctions delta introduites dans l'\'{e}tape 3 simplifient \'{e}norm\'{e}ment les calculs, et on n'a souvent m\^{e}me pas \`{a} les faire.
  \item Calculer le facteur de sym\'{e}trie pour chaque graphique utilisant~\eqref{cng} et le multiplier par le r\'{e}sultat obtenu \`{a} l'\'{e}tape 5.
\end{enumerate}

Notons que pour tout processus physique, il peut y avoir un tr\`{e}s grand nombre de sch\'{e}mas possibles en fonction du nombre
d'interaction envisag\'{e}s. Quand l'ordre de l'interaction augmente, le nombre et la complexit\'{e}
des sch\'{e}mas possibles augmente rapidement. Toutefois, la constante d'interaction est g\'{e}n\'{e}ralement une quantit\'{e} tr\`{e}s faible (dans
l'\'{e}lectrodynamique quantique, il est num\'{e}riquement \'{e}gale \`{a} environ 1/137), la petitesse de ce terme pour $n$ grand r\'{e}duit efficacement la probabilit\'{e} qu'un processus compliqu\'{e} se produise effectivement. Cela justifie l'utilisation de l'approche perturbatif.

\paragraph{Exemples.}

Pour voir comment ces r\`{e}gles fonctionnent, nous allons voir deux exemples. Pour le diagramme \vskip-30pt
\[\includegraphics[width=9cm]{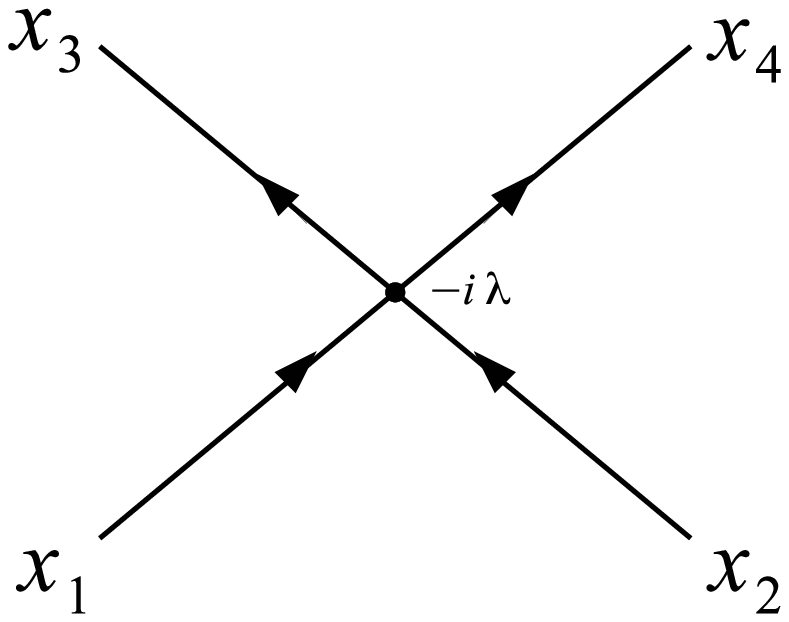} \vspace{-0.2cm}\]
Ici $n=1$ et, par cons\'{e}quent, $C=1$ et $M=1$, l'amplitude est donc simplement $-i \lambda$.

Maintenant, calculons l'amplitude pour le diagramme \`{a} l'ordre 2 suivant
\[\includegraphics[width=9cm]{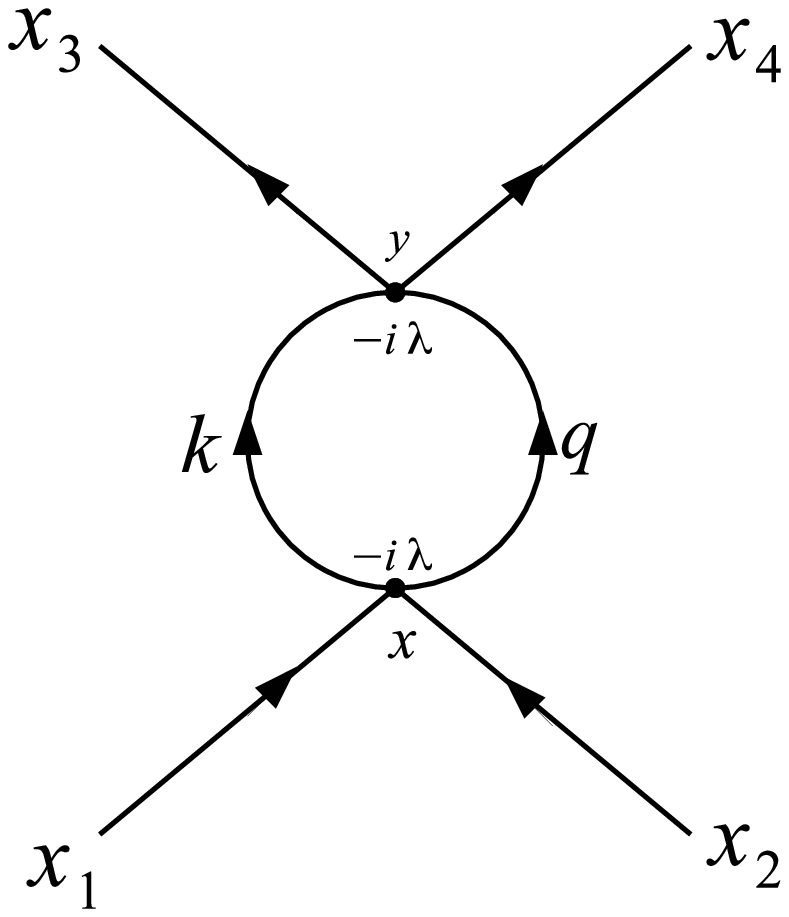} \vspace{-0.8cm}\] Nous avons deux points d'interaction qui contribuent avec le facteur $(-i \lambda)^2$. Nous avons deux lignes internes, donc nous avons \[\overline{Z}=(-i \lambda)^2 \iint \frac{\d^4 k}{(2\pi)^4} \frac{\d^4 q}{(2\pi)^4}\frac{i}{k^2-m^2}\frac{i}{q^2-m^2}(2 \pi)^4 \delta^4(k_1+k_2-k-q) (2 \pi)^4 \delta^4(k+q-k_3-k_4)\] simplifiant, on trouve \[\overline{Z}=\lambda^2 \int \frac{\d^4 k }{(2\pi)^4}\frac{1}{k^2-m^2}\frac{1}{(k_1+k_2-k)^2-m^2} (2\pi)^4\delta^4(k_1+k_2-k_3-k_4) \] Il y une fonction de Dirac r\'{e}siduelle qui traduit la conservation de la 4-impulsion total. On peut s'en d\'{e}barrasser. Pour ce diagramme, $s_{xx}=s_{xx}=0$, $s_{xy}=2$. Utilisant~\eqref{cng}, on a \[C=\frac{1}{2!}=\frac{1}{2}.\] L'amplitude est donc
\[\overline{Z}=\frac{1}{2}\lambda^2 \int \frac{\d^4 k}{(2\pi)^4} \frac{1}{(k^2-m^2)}\frac{1}{(k_1+k_2-k)^2-m^2}\]
Mais, est-ce que cette int\'{e}grale converge ? NON!!!, car pour $k \to \infty$, l'int\'{e}grand se comporte comme $\log k \to \infty$. C'est le c\'{e}l\`{e}bre probl\`{e}me de la divergence ultraviolette. La solution repose sur une acrobatie qui est vivement critiqu\'{e} pour son manque de ``rigueur math\'{e}matique'' selon les mots de Paul Dirac. Cette technique marche pourtant bien et donne des r\'{e}sultats satisfaisant; elle est connue sous le nom de renormalisation. Malheureusement le temps r\'{e}serv\'{e} \`{a} ce cours ne permet pas de s'\'{e}taler sur ce point.

\end{document}